\documentclass[12pt,preprint]{aastex}

\slugcomment{}

\shorttitle{Large-scale Alfv\'en waves in solar flares}
\shortauthors{Fletcher \& Hudson}
\begin{document}
\title{Impulsive phase flare energy transport by large-scale Alfv\'en waves and the electron acceleration problem}

\author{L.~Fletcher\altaffilmark{1}\email{lyndsay@astro.gla.ac.uk}}
\affil{Department of Physics and Astronomy, University of Glasgow, Glasgow G12 8QQ}
\author{H.~S. Hudson\email{hhudson@ssl.berkeley.edu}}
\affil{Space Sciences Laboratory, University of California, Berkeley, CA 94720}

\altaffiltext{1}{Carried out while a Visiting Researcher, Space Sciences Laboratory, University of 
California, Berkeley, CA}

\begin{abstract}
The impulsive phase of a solar flare marks the epoch
of rapid conversion of energy stored in the pre-flare coronal magnetic
field.  Hard X-ray observations imply that a substantial fraction of
flare energy released during the impulsive phase is converted to the
kinetic energy of mildly relativistic electrons (10-100~keV).  The
liberation of the magnetic free energy can occur as the coronal
magnetic field reconfigures and relaxes following reconnection. We
investigate a scenario in which products of the reconfiguration --
large-scale Alfv\'en wave pulses -- transport the energy and
magnetic-field changes rapidly through the corona to the lower
atmosphere.  This offers two possibilities for electron
acceleration. Firstly, in a coronal plasma with $\beta < m_e/m_p$, the
waves propagate as inertial Alfv\'en waves. In the presence of strong
spatial gradients, these generate field-aligned electric fields that
can accelerate electrons to energies on the order of 10~keV and above,
including by repeated interactions between electrons and
wavefronts.  Secondly, when they reflect and mode-convert in the
chromosphere, a cascade to high wavenumbers may develop.  This will
also accelerate electrons by turbulence, in a medium with a locally
high electron number density.  This concept, which bridges MHD-based
and particle-based views of a flare, provides an interpretation of the
recently-observed rapid variations of the line-of-sight component of
the photospheric magnetic field across the flare impulsive phase, and
offers solutions to some perplexing flare problems, such as the flare
``number problem'' of finding and resupplying sufficient electrons to
explain the impulsive-phase hard X-ray emission.
\end{abstract}


\section{Introduction}\label{sect:intro}

Strong chromospheric hard X-ray emission and strong UV and white-light
emission mark the impulsive phase of a solar flare. These signatures
are usually interpreted in terms of the well-known ``thick-target
model'' \citep{Brown71,1972SoPh...24..414H} in which fast electrons
lose energy in Coulomb collisions and ionizing collisions in the
chromosphere, heating and producing bremsstrahlung {\it en route}. The
inefficiency of the bremsstrahlung process in a cold thick target
implies that a large fraction of flare energy resides in these
electrons \citep{Kane71,Lin76,Holman03}, and calculations under the
assumptions of the thick-target model yield numbers on the order of
$10^{34} - 10^{37}$ electrons accelerated per second
\citep[e.g.][]{1997ApJ...491..939M,Holman03}.  Various strands of
evidence have led to the commonly-accepted idea that the particle
acceleration takes place in the solar corona, following which the
electrons propagate into the lower atmosphere where they heat, and
generate the observed hard X-ray radiation. Extensive theoretical work
over four decades (which we will not attempt to summarize here)  has
elucidated the basics and the specifics of numerous different coronal
acceleration mechanisms, in the electric fields present in current-sheets 
and X-lines/points generated by reconnection, in large- and small-scale 
plasma waves and turbulence, and at shocks.  Recent reviews can be 
found in \cite{2002SSRv1011A} or \cite{2003LNP...612..213L}, for example.
However, a  coronal acceleration site still
presents some problems for theory. The primary difficulty, especially
in the context of the high intensity of the energy deposition implied
not only by hard X-rays but also by UV and white-light continuum
observations \citep[e.g.][]{2007ApJ...656.1187F}, is the so-called
``number problem''  - the high total number of electrons required
compared to that available in the corona - and the  associated (and in
fact more problematical) supply and re-supply problems.

The thick-target model as normally understood requires intense
electron beams to transport the flare energy. We propose instead that
flare energy is transported by the Poynting flux of Alfv{\' e}n
waves. Since flare energy release implies large-scale restructuring of
the  coronal magnetic field (e.g. via reconnection) it is natural to
expect the excitation of such waves \citep{1982SoPh...80...99E}.  The
electron acceleration can then take place where the waves dissipate,
in the legs of the coronal loops or in the chromosphere itself.

The possibility of flare energy transport by Alfv{\'e}n waves has been
discussed before, for example by  \cite{1982SoPh...80...99E} in the
context of heating the temperature minimum region, and more  generally
by  \cite{1992ApJ...387..403M} and \cite{1994PASAu..11...25W} who
examined the  propagation of twist in a flare loop.  The present paper
seeks to restart the discussion of flare wave  energy transport, in
the light of recent solar observations and recent developments in
magnetospheric  physics, as well as because of the outstanding
theoretical issues with coronal  electron acceleration,  which have
been exacerbated by RHESSI, TRACE and other observations.

The main solar physics drivers for revisiting this idea are as
follows. Firstly,  recent microwave  (gyrosynchrotron) observations of
the corona above active regions demonstrate conclusively that
magnetic field strengths of several hundredths up to more than a
tenth of a Tesla (i.e. several 100s of  Gauss to kG) exist in the
cores of active regions, measured at heights up to  10,000 - 15,000~km
above  the photosphere. Coupled with reasonable coronal densities of
$10^{15}{\rm m}^{-3}$ these fields imply  Alfv{\'en} wave speeds well
above $10^4{\rm~km~s}^{-1}$, and correspondingly high Poynting fluxes. The
observational basis for these physical parameters described in some
detail in Section~\ref{sect:phys}.  Secondly, there is clear evidence
that substantial perturbations to the photospheric magnetic field (on
the order of 0.01 to 0.02~T) occur during solar flares. Field changes
in the low corona, on height scales  comparable to the horizontal
dimensions of active regions, must be of similar magnitude. This
strongly  suggests a violent perturbation to the magnetic field, at a
low level in the atmosphere, which is at least  qualitatively
consistent with a very energetic magnetic disturbance
 
In magnetospheric physics, electron acceleration in the parallel electric field that results from the propagation of
large-scale Alfv\'en waves and wave pulses in a non-idealised MHD fluid is a  promising prospect
for auroral electron acceleration, and also motivates us in this work.
In the magnetospheric/ionospheric context it was pointed out early on
that non-ideal effects arise from considering both the two-fluid nature of the plasma (i.e. treating electrons as a separate fluid, and including their inertia and thermal pressure) and also the particle aspects of the problem (e.g., the  finite ion
gyroradius). These lead to field-aligned electric fields, and the presence of such
dispersive Alfv\'en waves and their link to electron acceleration is
now well-established observationally
\citep[e.g.][]{2002JGRA.107hSMP24W}.  \cite{2002GeoRL..29k..30C} have
demonstrated that the value of the energy flux carried by auroral
electrons is similar to the Poynting flux of low frequency Alfv\'en
oscillations of the magnetospheric field. Debates persist about the
precise mechanism for generating the electric fields that accelerate
auroral electrons \citep[e.g.,][]{2000SSRv...92..423S}, but the
inertial  Alfv\'en wave (see Section~\ref{sect:iaw}) is a strong
candidate. This may also have a role to play in the  case of flares,
although  the solar and magnetospheric cases of course represent very
different  parameter regimes. We demonstrate in Section~\ref{sect:iaw} that the inertial Alfv\'en wave mode is also the appropriate one to consider for flare parameters. The critical factor in determining the parallel electric field that can be generated is the spatial scale of perpendicular structuring of the magnetic field compared to the electron inertial length, and - as we describe - observations at ever higher resolution are showing finer and finer magnetic field stucturing. 

We note also that electron acceleration
by non-ideal MHD waves is also making its way into the discussion of
coronal heating.  \cite{2006PhRvL..96q5003S} and
\cite{2007PhRvL..98d9502S} claim that dispersive Alfv\'en waves driven
by photospheric turbulence lead to parallel electric fields and
electron heating, and \cite{2006A&A...455.1073T} finds the generation
of a parallel electric field and runaway electron heating when an
initially ideal (non-dispersive)  but non-linear Alfv\'en wave couples
to dissipative modes when it is launched into a corona with transverse
density structure. Our considerations are somewhat different  from this idea, 
in that our inertial Alfv\'en wave is  dispersive from the start. This does not preclude also the kind of mode coupling discussed by  \cite{2006A&A...455.1073T}; instead this would be an additional energy loss term which will require further study in the future.

We first describe the proposed mechanism in
Section~\ref{sect:mechanism}, including a detailed  description of the
observations that motivate us. The hard X-ray observations, as
confirmed by RHESSI,  require powerful electron acceleration, and in
Section~\ref{sect:acceleration} we discuss possibilities for  this in
the framework of the wave transport model.
Section~\ref{sect:energetics} then considers the overall implications for  
flare energetics.

\section{The proposed mechanism}\label{sect:mechanism}

\subsection{The waves}

\subsubsection{Wave source} 
The release of stored magnetic energy
requires a re-structuring of the field, for example as envisioned in
large-scale magnetic reconnection,  However the amount of magnetic
free energy that can be {\it dissipated} within the reconnection
region itself -- current sheet, X-point or 3-D null -- is restricted,
given its small dimensions and the short flare time-scale.  The more
important release of free energy occurs in the large-scale
`convulsion' as the newly--reconnected magnetic field relaxes from its
pre-flare stressed state.  Where they detach from the coronal
current-sheet or null structure but are still stressed, these magnetic
field lines will be highly distorted from a potential configuration,
with a locally high tension force.  We know observationally that the
impulsive energy release occurs in a highly-stressed magnetic field,
with large fluctuations on time  scales ranging down to a fraction of
a second  \citep[e.g.,][]{1985SoPh..100..465D}. This implies irregular and time-varying structures in a
three-dimensional reconnection flow. Thus Petschek
reconnection, which is essentially steady-state,  cannot properly
describe it.  

The rapid restructuring of the field implies an energy flow describable
in terms of MHD wave propagation, and we  infer that it will create a
complicated mixture of fast-mode, slow-mode, and Alfv\'en-mode propagating wave
pulses in the  adjacent plasma. For example, flare loop `shrinkage'
\citep[e.g.,][]{1996ApJ...459..330F}  identifiable with the MHD  fast
mode is a simple and well-known illustration of this idea, as are the
slow-mode shocks of Petschek  fast reconnection. MHD modeling of 
three-dimensional reconnection is at an early stage,  but in three
dimensions a torsional component will in general also exist,
particularly in a reconnecting twisted field
\citep{1982SoPh...80...99E}.  Indeed, in-situ observations of
reconnection in the solar wind \citep {2005JGRA..11001107G} show
Alfv\'en waves propagating along just-reconnected field lines, and
the MHD simulations of \cite{2006ApJ...642.1177L} demonstrate a
post-reconnection state of  initially untwisted flux tubes in which
field-line kinks propagate away at close to the Alfv\'en speed. Since we require to deposit flare energy in the flare footpoints, we require a wave mode that propagates along the magnetic field - either the Alfv\'en mode or the slow mode. However, the slow mode speed is too low to explain the observed footpoint simultaneity unless we have extremely symmetric propagation from exactly half way between the footpoints. For the same reason of low speed, neither can it explain the require high energy flux (see Section 4).  Thus, we work under the assumption of an Alfv\'enic disturbance carrying energy
along the post-reconnection field.

\subsubsection{Wave development}\label{sect:develop} 
\begin{figure}
\plotone{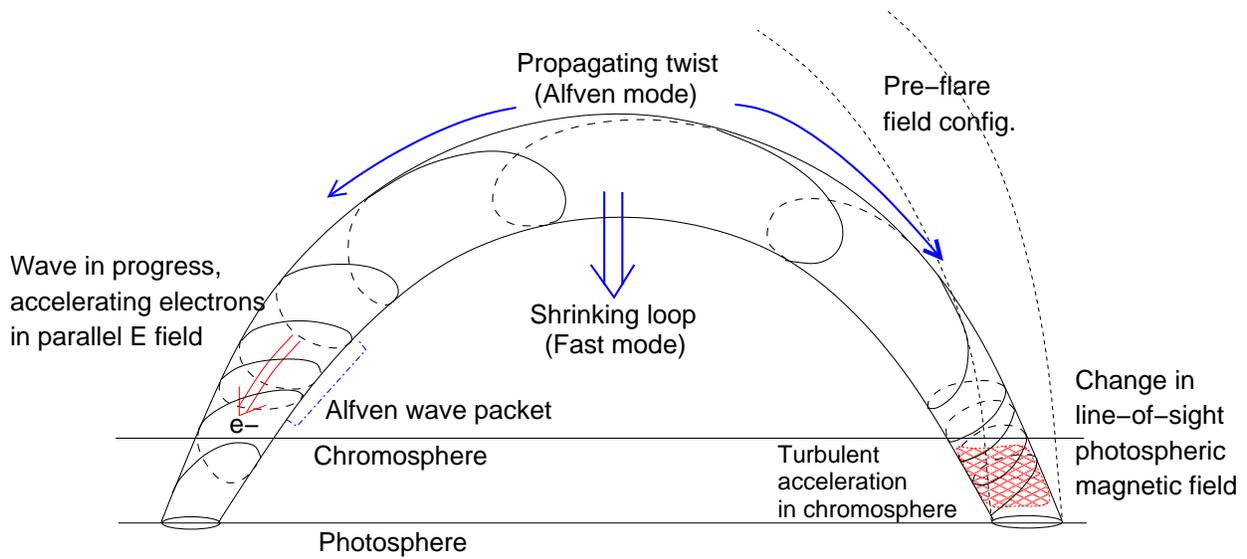}
\caption{The reconfiguring coronal field 
  launches a torsional Alfv\'en wave pulse through the corona
  and into the chromosphere, as well as a fast-mode wave pulse. The Alfv\'en
  wave, which propagates in the inertial regime, can lead to electron
  acceleration in the corona. That fraction of the Alfv\'en wave energy that
  survives into the chromosphere can also lead to stochastic
  acceleration there. The wave will be partially reflected from the
  steep gradients in the chromosphere (not shown) and re-enter the
  corona.}
\label{fig:cartoon}
\end{figure}

We sketch our
overall view of a post-reconnection loop and the processes taking
place in it in Figure~\ref{fig:cartoon}. The perturbation in 3D takes
the form of fast-mode and Alfv{\' e}n-mode wave pulses
\citep{1982SoPh...80...99E}; the group velocity of the Alfv{\' e}n
mode is parallel to the magnetic field {\bf B}, so this component of
the energy propagates directly to the footpoints as shown in the
cartoon. In the MHD view the propagation speed is just the Alfv\'en
speed \citep[in a kinetic treatment][also
recovered this result]{1979JGR....84.7239G}.

The wave spectrum will be determined by the largely-unknown geometry
of the energy release. It is  likely that the Alfv\'en wave will take
the form of a short-wavelength propagating pulse - a wavefront -  with
parallel wavelength much smaller than the length $L$ of a
just-reconnected loop. The perpendicular wavelengths would be much
smaller than the  loop length, as dictated by the reconnection rate
and its fluctuations.

Under appropriate conditions (Section~\ref{sect:phys}) the Alfv\'enic
perturbation will propagate rapidly through the coronal field to the
chromosphere without significantly cascading to smaller scales.  This
is different from (but complementary to) the view of
\cite{1994ApJ...425..856L} and \cite{1997ApJ...491..939M}, in which
the large-scale fast-mode waves formed by reconnection are assumed to
cascade rapidly to short-wavelength turbulence within the coronal
loop, leading eventually to the Fermi acceleration of electrons in
high-frequency turbulence directly in the corona.  For the ducted
Alfv{\' e}n mode it has been shown \citep[e.g][]{PhysRevE.57.7111,
2005ApJS..156..265C} that  a cascade will not develop immediately.
Therefore in the situation we envisage, the Alfv\'enic perturbation
will move from corona to chromosphere along a strong guide field
without driving a cascade, at least in the initial pass. The wave
energy will be strongly ducted towards the chromosphere. 

If some fraction of the wave energy is reflected at the chromosphere,
so that counter-moving waves are present in the corona, then a cascade
may occur.  However, even then, \cite{PhysRevE.57.7111} demonstrate
using reduced MHD simulations that the cascade to high parallel
wavenumbers is inhibited, and an exponentially-decaying rather than a
power-law spectrum will be formed, while the cascade to high {\it
perpendicular} wave numbers proceeds independently. 

On arriving at the chromosphere and photosphere the wave propagation
will become more complicated, with transmission, reflection and
damping all playing a role.  The waves will undergo different kinds of
damping, including -- in the temperature-minimum region -- significant
ion-neutral damping. The line-tied boundary conditions at the
photosphere mean that the purely Alfv\'en disturbance will not survive
as such but instead, as demonstrated by \cite{1994A&A...286..275G}, a
reflected wave spectrum with hybrid characteristics will be generated,
and some fast-mode-like components will arise, particularly in the
presence of chromospheric small-scale structuring and flows. Being
compressional, these fast-mode-like waves can be locally damped by
other mechanisms, and offer also the possibility for a turbulent
cascade development in the chromosphere, analogous to that proposed
by  \cite{1994ApJ...425..856L} for  coronal acceleration.  The
analysis of \cite{1994A&A...286..275G} suggested that any reflected
waves that do re-emerge into the corona would have a mostly torsional
(Alfv\'enic) character.

\subsection{The particles}

The hard X-ray observations unambiguously require powerful  electron
acceleration.  How can this arise from energy transported in the
Poynting flux of Alfv{\' e}n waves?  We discuss possible mechanisms in
Section~\ref{sect:acceleration} and  briefly comment here on the
particle behavior in the context of Figure~\ref{fig:cartoon}.  In the
new scenario the acceleration of the energetically important
10-100~keV electrons either takes place in the legs of the flaring
loops, or actually in their footpoint regions.

Alfv\'enic perturbations propagating in the limit $\beta <
\rm{m_e/m_p} $ (the inertial Alfv\'en wave limit) lead to a parallel
electric field $E_{\parallel}$.  For a wave traveling downwards,
electron inertia produces an upwards $E_{\parallel}$.  A fraction of
the electrons are resonantly accelerated in this field, in a process
that can be thought of as an encounter with a moving mirror
\citep{1994JGR....9911095K,chaston06}, with the electrons reflecting
from the traveling perturbation front and accelerating to twice the
Alfv\'en speed $v_A$.  In the conditions we envisage, where the
Alfv\'en speed is on the order of 0.1-0.3~$c$ (see
Section~\ref{sect:phys}), this corresponds to an `Alfv\'en energy'
$( = {1\over 2} m_e v_A^2)$ in the few to tens of keV range.  Multiple
reflections of the electron between the wave front and magnetic mirror
formed by the converging chromospheric magnetic field may occur, each
reflection from the wave front increasing the electron speed by $2
v_{A}$ in first-order Fermi acceleration.

As mentioned in Section~\ref{sect:develop} a turbulent wave spectrum
may be generated in the footpoint regions.  In the chromosphere, the
damping of this spectrum will broadly-speaking result in plasma
heating, since the electron-electron thermalization times are very
short. However, an essentially collisionless tail of fast electrons
can be accelerated by Fermi processes, as in the case of coronal
stochastic acceleration. The question is how large that tail may be. 
We
discuss this in Section~\ref{sect:turb} but note here that stochastic
acceleration can take place in a collisional environment
\citep[e.g.,][]{1992ApJ...398..350H}. The particular advantages
offered by chromospheric acceleration are firstly a high ambient
electron density (compared to the corona), possibly easing the number
and resupply problems, and secondly - as pointed out by Brown (2006,
private communication) and \cite{2006GMS...165..157M} - , the
requirement on the total number of accelerated electrons implied by
their hard X-ray signature is reduced if the accelerator acts on them
at the same time as they radiate bremsstrahlung emission, which would
be satisfied in a chromospheric accelerator. (This advantage is
analogous to the increased bremsstrahlung efficiency that pertains in
a thermal model for flare hard X-rays, where the radiating electrons
are continually re-boosted by interactions with a hot rather than a
cold target \cite[e.g.][]{1979ApJ...232..582S}.)

This overall scenario also provides a mechanisms for some accelerated
electrons to appear in the corona.  This is important because of the
extensive observational evidence for coronal non-thermal  electrons,
e.g. via the microwave spectrum, or low energy hard X-rays.  Any
reflected component of the inertial Alfv\'en wave pulse produces a reversed
electric field, which can draw chromospheric electrons back into the
corona.  Furthermore, coronal electron acceleration by the cascade of
fast-mode turbulence - as proposed by \cite{1994ApJ...425..856L} - may
operate alongside the Alfv\'enic transport, as both wave types will be 
generated by the reconnection process.

\subsection{Physical Parameters\label{sect:phys}}

The properties of the Alfv\'en waves, and the magnitude of the
parallel electric fields they generate, depend critically on the plasma
parameters; density, electron and ion temperatures, magnetic field
strength and length scales. We review the relevant observations here.

\bigskip
\noindent {\bf Magnetic field strength:} 

It is notoriously difficult to determine the strength of the coronal
magnetic field, or to calculate it by extrapolations from a given
boundary. However, in solar flares and in the cores of active regions,
where the magnetic field is strong,  simple geometrical arguments
point to intense fields in the low corona.  A large sunspot may have a
size scale of some 3~$\times$~10$^4$~km, an umbral field of  a few
$\times$~0.1~T, and an outer penumbral field of 0.08 to 0.17~T
\citep{2003A&ARv..11..153S}.  For the dominant dipole term of a
multipole expansion of this photospheric source structure, we would
expect comparable coronal field intensities, at heights in the
vicinity of the spot comparable to the spot extent.

Direct measurement of the strength of (strong) coronal magnetic fields
is  also possible via the microwave gyrosynchrotron spectrum generated
by  fast electrons. Very Large Array radio observations of active
regions show emission consistent with {\it average} active  region
coronal field strengths of a few~$\times$~0.01~T
\citep{1998ApJ...501..853L} at a height of 10,000~km above the
photosphere. In the corona above sunspots, even stronger  fields
have been measured
\citep{1991ApJ...366L..43W,1994PASJ...46L..17S,2002ApJ...574..453B,
2006PASJ...58...11V,2006ApJ...641L..69B}. For example,  using VLA and
SOHO data, \cite{2002ApJ...574..453B} deduce field strengths in excess
of 0.1~T at heights of 10,000~km above the photosphere over a sunspot
on the disk, and for a substantial area around it.  Limb observations,
with less confusion in the dependence of the field strength on
altitude \citep{2006ApJ...641L..69B}, also give these values.  Based
upon these observations, we can reasonably expect field strengths of a
few~$\times$~0.01~T at heights of 10,000~km above sunspot or  strong
plage regions, and since flare ribbons also penetrate into sunspot
umbrae, low-coronal fields $> 0.1\ T$ are certainly not out of the
question. These magnetic field strengths are substantially higher than
the values inferred from coronal seismology, however the coronal 
seismology technique has only been applied so far to large active-region
loops \citep[e.g.,][]{2001A&A...372L..53N}.

The height of 10,000~km at which these strong fields are observed is 
also consistent with the height of loops
involved in flares, based on their typical HXR footpoint separations of
typically a few tens of arcseconds. There are not to our knowledge
any  statistical studies of this, but numerous examples can be seen in e.g.
\cite{1994PhDT.......335S}, \cite{2005ApJ...630..561B}, \cite{2006A&A...456..751B}, \cite{2007ApJ...656.1187F}.  A
typical separation value of  30" or 20,000~km corresponds to a
semicircular loop with apex height of 10,000~km. 

\bigskip
\noindent {\bf Density:} With the exception of coronal soft X-ray
`knots' \citep[e.g.][]{1995ApJ...440..370D} and rare observations of
dense coronal loop flares which show negligible footpoint emission
\citep[e.g.][]{2004ApJ...603L.117V}, the coronal density before and
early in a flare is fairly low.  Several studies have sought pre-flare
signatures of the bright flare loops but the general result is that in
most cases no feature visible in soft X-rays matches the flare loops
that form after the impulsive phase
\citep{1996SoPh..165..169F,1998SoPh..183..339F}. This suggests that
the energy release takes place in regions of yet lower density than
the average active-region corona. Normal active-region loop densities
are  on the order of $1-3 \times 10^{15}{\rm cm}^{-3}$
\citep{2003A&A...406.1089D}  and even post-flare arcade loop
measurements  \citep{2000A&A...355..769V,2003ApJ...582..506L} are a
few $\times 10^{15}{\rm m^{-3}}$, which might reasonably be taken as
an upper limit for the pre-flare density in the flare region.  In the
study of a sunspot magnetic field mentioned above,
\cite{2002ApJ...574..453B} estimated  plasma densities at a few
$\times 10^{14}{\rm m}^{-3}$ to $10^{15}{\rm m}^{-3}$ in the
essentially `empty' corona above a sunspot.  Finally,
\cite{1999ApJ...520L.135F} find upper-transition region densities of
$2-5 \times 10^{15}{\rm m}^{-3}$ in the cores of active regions, again implying a lower density for the overlying hotter
corona. 
%It should also be noted that line-ratio diagnostics applied in
%an inhomogeneous plasma are biased towards the higher densities
%\citep[e.g.,][]{1989A&A...224..328A}.  
Taken together, these various
strands of evidence imply that pre-flare coronal densities on the
order of $10^{15}{\rm m}^{-3}$ or possibly smaller are common, and in
many cases we have only upper limits.

\bigskip
\noindent {\bf Alfv{\' e}n speed:}
If we take a magnetic field strength of 0.05~T and a proton number
density $n_p = 10^{15}$~m$^{-3}$,  in a fully-ionized hydrogen plasma
the Alfv\'en speed is $3.5~\times$~10$^{4}$~km~s$^{-1}$. Higher values
of $\left| \mathbf{B} \right|$ or lower values of $n_p$ are also
possible, so $v_A$ could thus be a few~$\times$~0.1~c.  These may seem
like extreme values given that the `canonical' coronal value often
discussed is on the order of 10$^{3}$~km~s$^{-1}$, and that fast
coronal mass ejections -- presumably ejected at some fraction of the
local Alfv\'en speed -- travel at around 3,000~${\rm km\;s^{-1}}$
above a couple of solar radii.  However the measurements, and our
considerations, refer to the low corona, where the bulk of the
magnetic energy resides, in highly-stressed, compact fields.  Note
that since $(v_A/c)^2 << 1$ the wave can still be described
non-relativistically, and the displacement current may still be
neglected, allowing an MHD description. 

\begin{figure}
\plotone{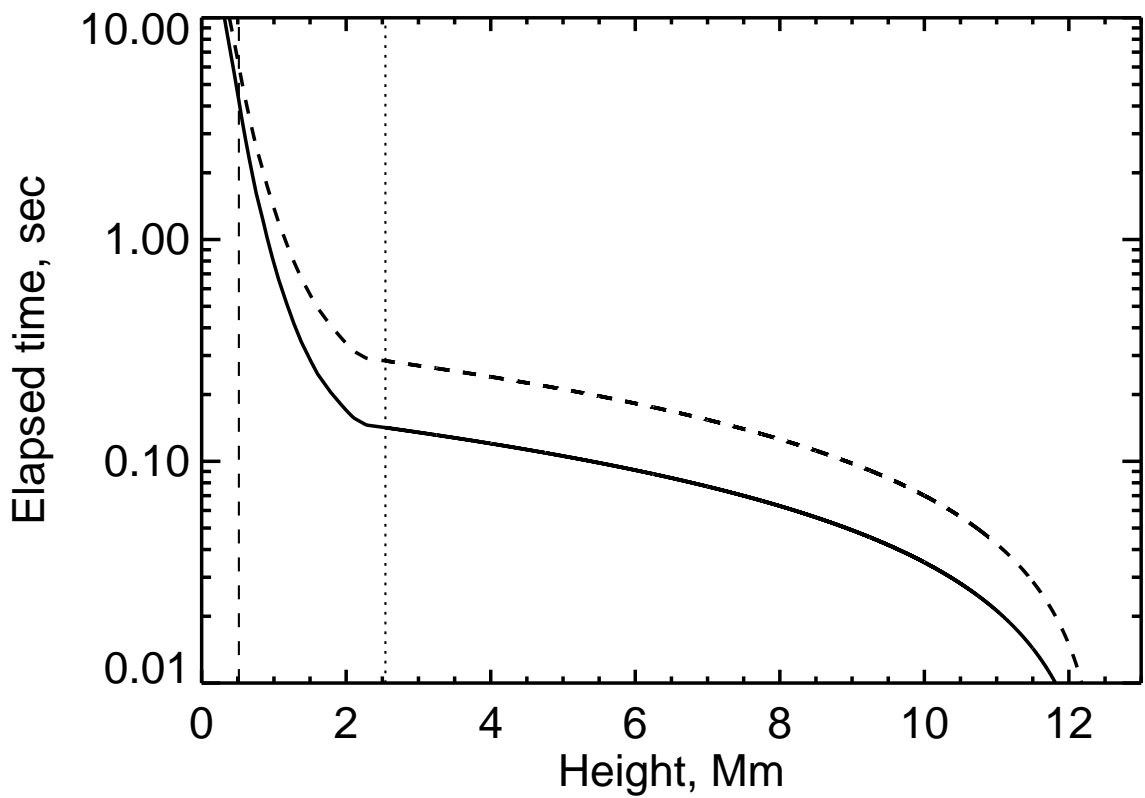}
\caption{Propagation time at the Alfv\'en speed from loop top to a
  given height. The vertical dashed line indicates the temperature
  minimum, and the vertical dotted the top, of the VAL-C atmospheric
  model. This model is extended into the corona with a semicircular
  loop of coronal half-length 10,000~km and a density scale height
  given by the temperature assumed for the base of the corona, $10^6$~K. 
  The dashed curve shows a coronal field of 0.05 T, extended
  through the atmosphere with $\alpha$ = 0.052 (see text); the solid
  curve the more extreme case of 0.1~T with $\alpha$ = 0.0. }
\label{fig:traveltime}
\end{figure}

Assuming a loop half-length of $10^7$~m, the propagation time of such a
wave into the chromosphere from a coronal launch site is a few tenths
of a second at most.  This is shown in Figure~\ref{fig:traveltime} for
a hydrostatic corona  at $T=10^6$~K matched to the top of the of the
VAL-C chromospheric model \citep{1981ApJS...45..635V}, and using the
chromospheric magnetic scaling of \cite{1983ApJ...264..648Z}, i,e.,
$\left| \mathbf{B} \right| \propto P_{g}^{\alpha}$, where $P_{g}$ is
the gas pressure. The parameter $\alpha$ has been chosen to give a
field strength at the photosphere of 0.2~T. The propagation time
obtained is adequate to explain the observed timescales of hard X-ray
emission as well as the simultaneity of hard X-ray footpoints
\citep{1994PhDT.......335S}, an argument often advanced in favor of
energy transport by energetic electrons accelerated in the corona and
precipitating at the footpoints. The commonly-observed pattern of
slower non-thermal variations in the later phase of a solar flare may
result from the increase of coronal densities and decrease in the
strength of the reconnecting fields in this phase, and thus reduced
Alfv\'en speeds.

\bigskip
\noindent {\bf Photospheric magnetic perturbations:} 
The observations of non-reversible changes to the line-of-sight
magnetic field at the photospheric level mentioned in
Section~\ref{sect:intro} lend credence to our supposition that strong
perturbations to the magnetic field are present throughout the
atmosphere. For example \cite{1999ApJ...525L..61C} and
\cite{2001ApJ...550L.105K} observed such changes in ground-based and
SOHO/MDI data respectively. \cite{2005ApJ...635..647S}, using
simultaneous SOHO/MDI and GONG magnetogram data, observe permanent
line-of-sight photospheric magnetic changes (0.01-0.02 T) to be 
``ubiquitous features'' of X-class flares at least. The changes are observed 
to be roughly co-spatial with the flare
ribbons and occur rapidly, on timescales of minutes. They are
therefore too fast to be due to Alfv\'enic perturbations propagating
upwards from the sub-photospheric region. Rather, it is as if the
magnetic field at the photospheric level is `jerked' by the overlying
magnetic field as it restructures in the corona, with both a twisting
component and a loop retraction. 
The fact that we see a distortion to the photospheric magnetic field indicates
that there is substantial wave energy transmitted to low levels in the
atmosphere, although with present line-of-sight observations we cannot
distinguish between components corresponding to twisting and retracting.

\bigskip
\noindent {\bf Transverse magnetic structuring:} 
As will become apparent in Section~\ref{sect:damping}, the transverse
scale of magnetic structure is a vital parameter in our calculations,
but observations are strongly limited by instrumentation. We do have
observed upper limits to the transverse structuring of the
chromospheric magnetic field in the quiet sun: in recent observations
using the Swedish Vacuum Solar Telescope, \cite{2004A&A...428..613B}
report that magnetic elements seen in the G-band (the photosphere)
appear unresolved at 70~km spatial resolution.  We may expect that
transverse photospheric structuring on still smaller scales  may be
present.  A lower limit to the transverse scales would be the
ion inertial  length, in the range 10$^{-2}$--1~km at the
transition region interface.

\begin{figure}
\centering
\begin{vbox}{
\plottwo{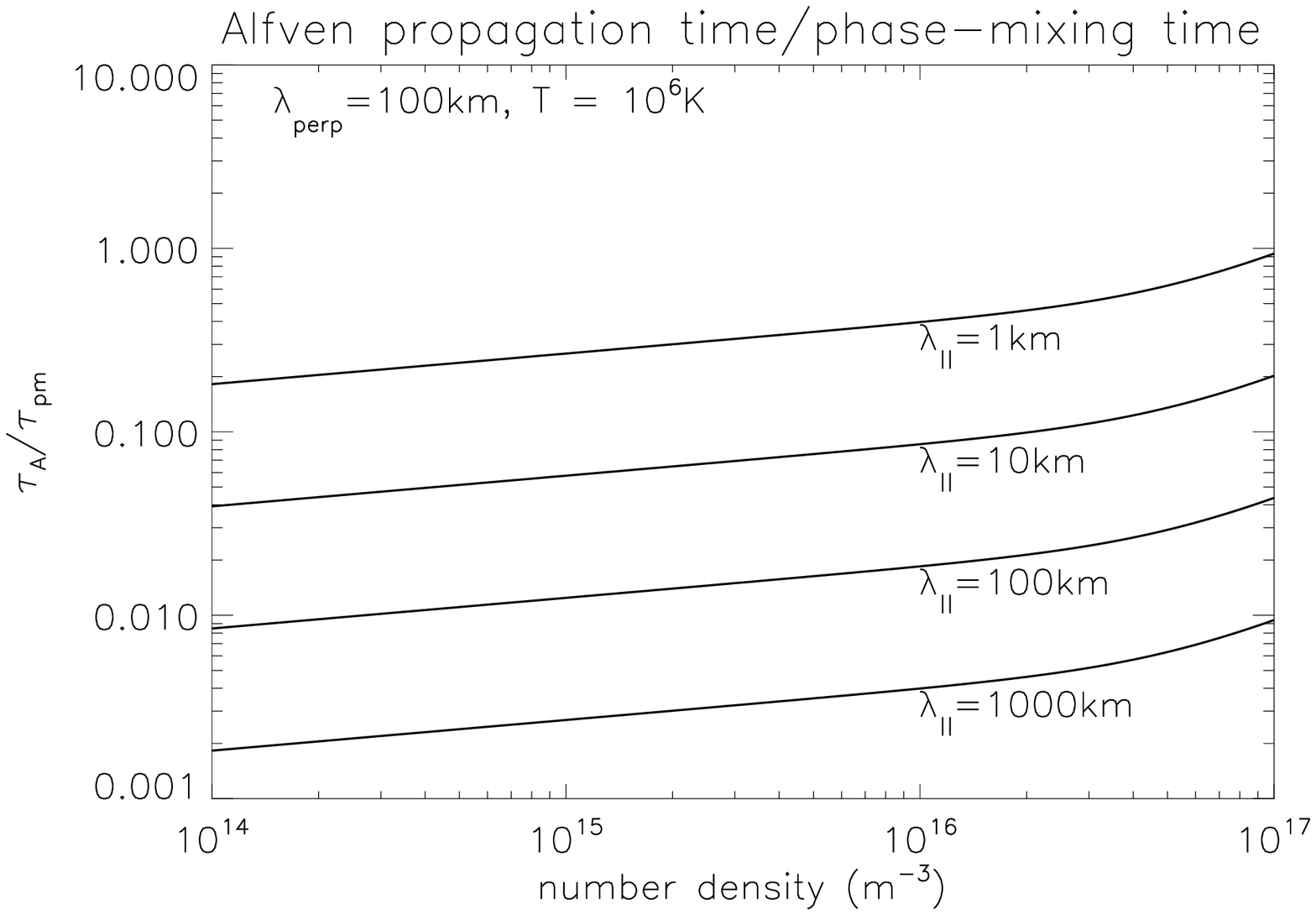}{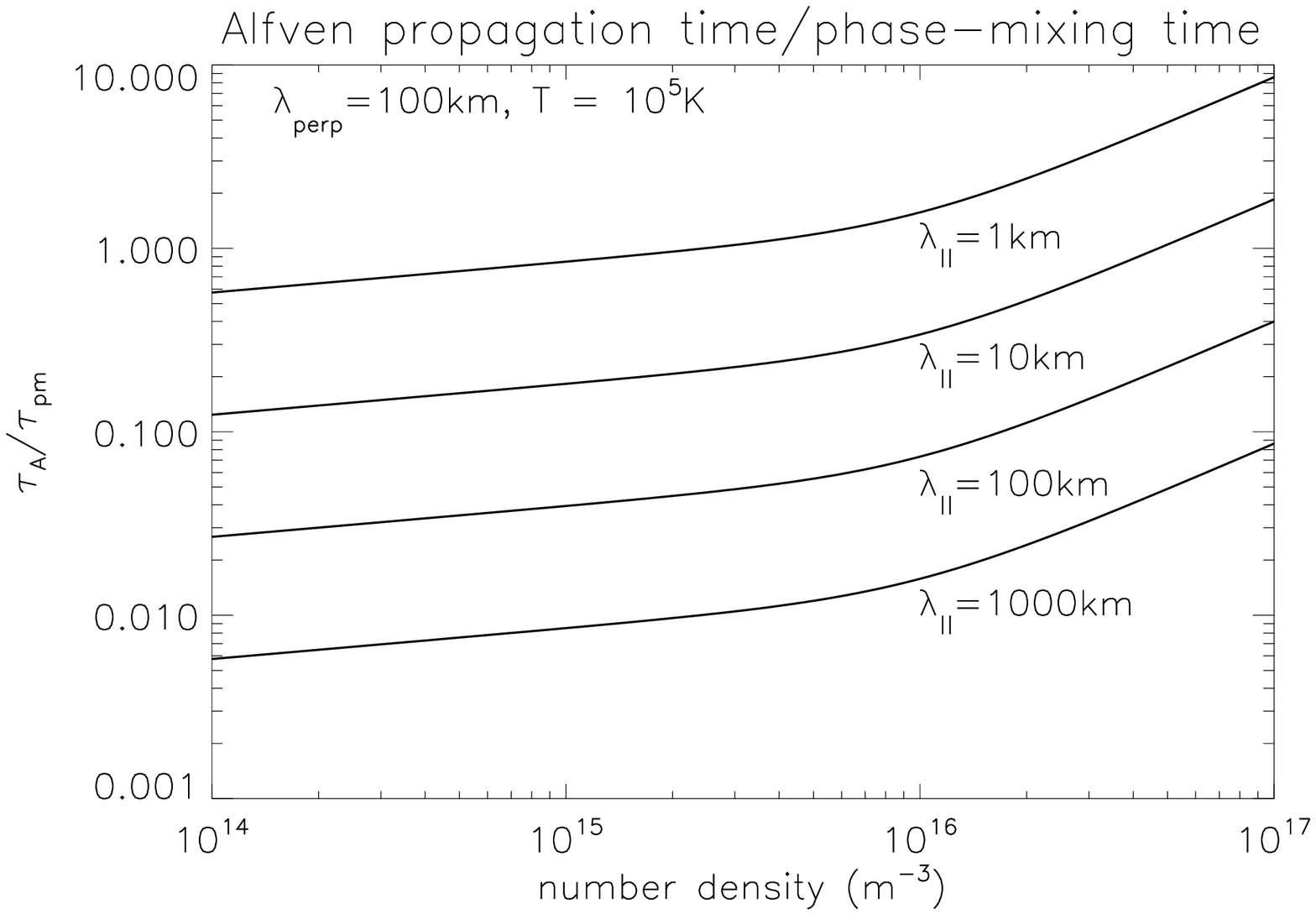}
\plottwo{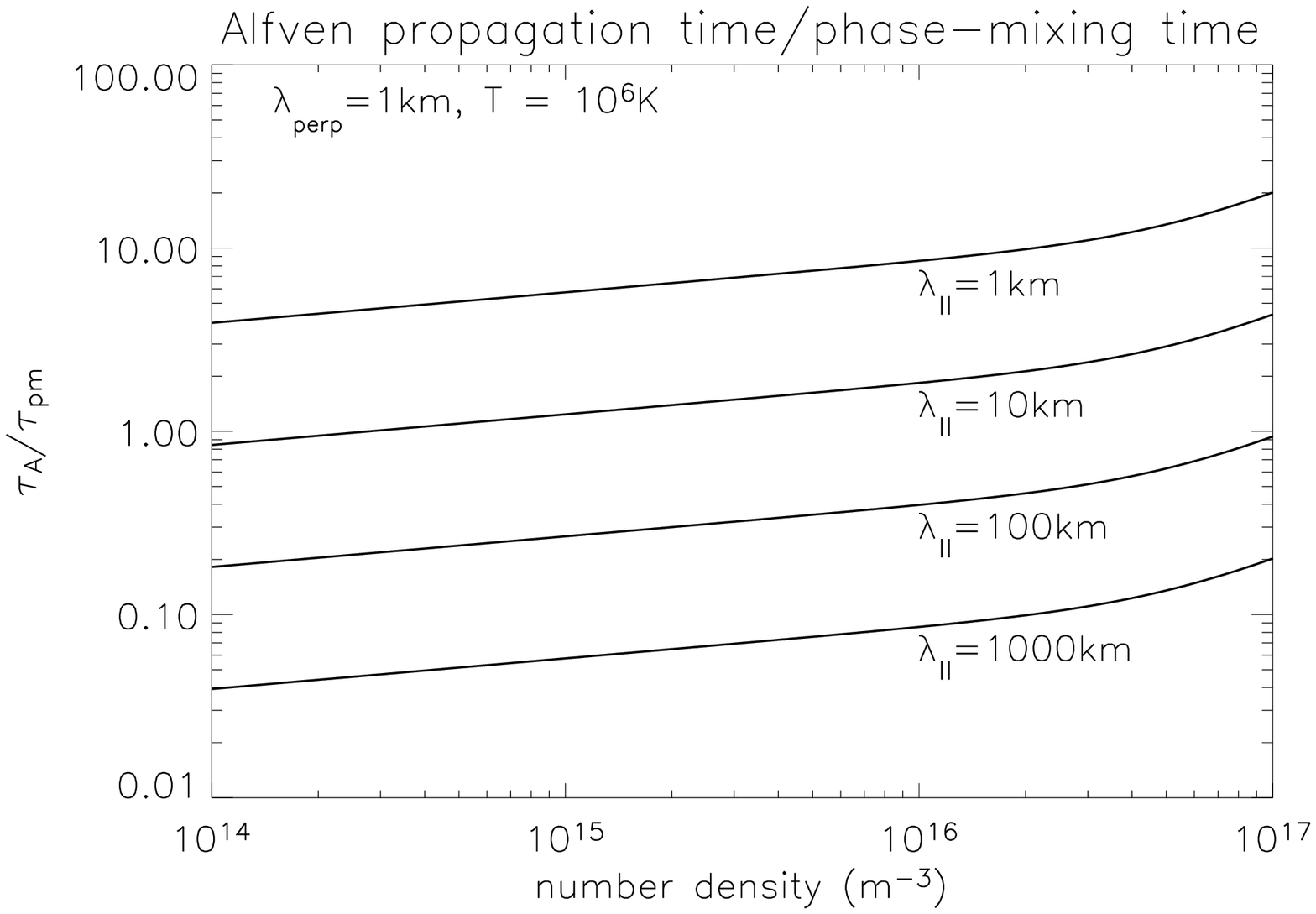}{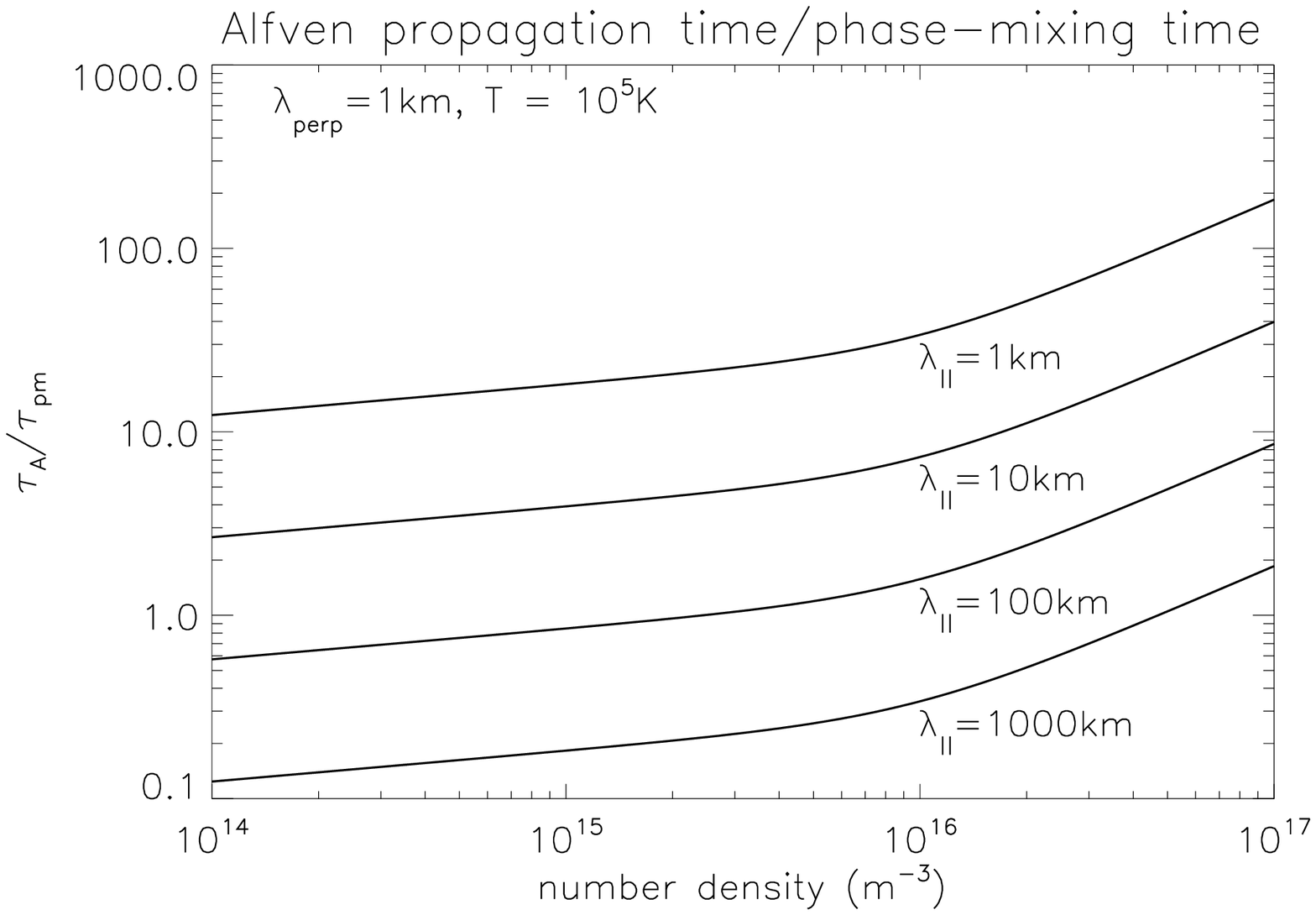}
}
\end{vbox}
\caption{The ratio of the Alfv\'en propagation time along a 10$^{4}$~km
  loop to the damping time by phase mixing, for a different wavelength
  perturbations in a coronal field of 0.05~T, a temperature of
  $10^6$~K (left hand panel) and $10^5$~K (right hand panel). The
  perpendicular wavelength of the perturbation is 10$^{2}$~km (upper row)
  and 1~km (lower row).}
\label{fig:phase}
\end{figure}

\subsection{Wave passage through the corona}\label{sect:damping}

We establish here that the coronal Alfv\'en wave pulses can traverse
the corona and arrive at the chromosphere without significant viscous
or resistive damping. In a corona with strong non-uniformities
perpendicular to the field, the damping of Alfv\'en waves is by phase
mixing \citep[e.g.,][]{2000SoPh..193..139R}.  The damping time is
given by his Equation~22, expressed here in terms of the wavelength:
\begin{equation}\label{eq:phasem}
\tau_{pm}= \left({6 \lambda_\parallel^2 \lambda_\perp^2}\over{4 \nu
    \pi^2  v_a^2}\right)^{1/3}
\end{equation} 
where $\lambda_\parallel$ and $\lambda_\perp$ are the parallel and
perpendicular wavelengths respectively. Under most conditions the
viscosity $\nu$ is the plasma shear viscosity, $\nu_s$, which is the kinematic
viscosity multiplied by $(\omega_i
\tau_i)^{-2}$ \citep{1983A&A...117..220H} where $\omega_i$ is the ion
gyrofrequency and $\tau_i$ the ion collision time. In circumstances
where this factor is much less than unity, Joule dissipation will
dominate, and the viscosity will be given by the magnetic diffusivity,
\begin{equation}
\nu_m = {1\over(\mu_o \sigma)},
\end{equation}
with $\sigma$ the Spitzer conductivity. The total viscosity we use in
Eq.~\ref{eq:phasem} is the sum of the shear and the Joule viscosity.
Figure~\ref{fig:phase} compares the phase-mixing time scale to
the  Alfv\'en propagation time $\tau_A$ along the coronal part of the 
loop, This shows that
$\tau_A/\tau_{pm} < 1$ for perturbations with parallel wavelengths of
more than a few tens of~km propagating in a coronal density of $\sim
10^{15}{\rm m}^{-3}$.  However, wave energy may be lost in accelerating 
particles, as we describe in the next section.

\section{Electron acceleration in the context of energy transport by Alfv\'en wave pulses}
\label{sect:acceleration}

If the wave energy is transported by ducted Alfv\'en wave pulses as we
suggest, there are several possibilities for electron acceleration; we
consider three,  most closely related to the wave nature of the
transport mechanism.  Firstly, in a hot, tenuous, strongly magnetized
coronal plasma, it may be possible to accelerate electrons directly in
the corona, in the parallel electric field generated by a dispersive
Alfv\'en wave pulse (Sections~\ref{sect:iaw} and ~\ref{sect:field}).
Secondly, associated with this is the possibility that the electrons,
accelerated ahead of the wavefront, mirror in the converging solar
magnetic field and return for repeated interactions with the wave
(Section~\ref{sect:fermi1}).  This comprises a first-order Fermi
acceleration process.  Thirdly, the wave energy can be dissipated in
or near the chromosphere  in a turbulent cascade which accelerates
electrons stochastically (Section~\ref{sect:turb}) and we discuss
separately the two primary models for turbulent electron acceleration;
stochastic resonant acceleration in high frequency whistler
turbulence, and transit-time acceleration in lower-frequency MHD
turbulence.  We consider first the acceleration by inertial Alfv\'en
waves.

\subsection{Inertial Alfv\'en waves}\label{sect:iaw}
In ideal MHD, the (massless) negative charge
carriers respond instantaneously to any parallel electric field
generated by the Alfv\'enic perturbation, shorting it out so that no
$E_\parallel$ exists.  An ideal MHD wave includes an $E_{\perp}$, but
this does not accelerate particles.  However, in a real plasma,
the electrons have (i) a finite mass and therefore inertia, and (ii) a
finite thermal speed and therefore a pressure.  Both of these
properties make parallel electric fields possible, which lead to the
dissipation of the wave energy by electron energization. The 
importance of electron inertia in generating parallel electric
fields in the magnetosphere/ionosphere was first discussed by
\cite{1979JGR....84.7239G}. 

We follow here the definitions of \cite{2000SSRv...92..423S}, who give
an overview of dispersive Alfv\'en waves. An inertial Alfv\'en wave
(IAW) results if the electron thermal speed is smaller than or
comparable to the Alfv\'en speed. The electric field is due to the
finite inertia of the electrons, which cannot respond instantaneously
to the wave perturbation. (If the electron thermal speed exceeds the
Alfv\'en speed, but the electron pressure gradient is important, then
wave is termed a kinetic Alfv\'en wave, or KAW). Alternatively, the
conditions correspond to an IAW if $\beta \le m_e/m_p$ (and a KAW if
$m_e/m_p \le \beta \le 1$).

The plasma $\beta$ is:
\begin{equation}\label{eq:beta}
\beta={2 \mu (n_p+n_h) k_B T \mu_o \over{\left|\mathbf{B}\right|^2}} 
\end{equation}
where $k_B$ is Boltzmann's constant, $\mu_o$ the permeability of the
vacuum, $T$ is the temperature (we assume that the electron and ion
temperatures are equal), $\mu$ the mean molecular weight, $n_p$ the
proton number density and $n_h$ the neutral hydrogen number density.
Although the neutral hydrogen does not respond directly to the
Alfv\'enic disturbance, it is strongly collisionally coupled to the ion
component \citep[e.g.][]{2001ApJ...558..859D} and thus modifies the Alfv\'en
speed in the lower atmosphere. It also provides a mechanism for damping
the wave in the lower atmosphere, which will locally heat the
chromospheric plasma. Taking a mean molecular weight of 0.6, and
assuming a completely ionized target of density $n_e = n_{15}\times
10^{15}{\rm m^{-3}}$ and temperature $T = T_6\times 10^6$~K, we have
\begin{equation}\label{eq:beta2}
\beta={2\times 10^{-8} {n_{15} T_6}\over{\left|\mathbf{B}\right| ^2}}, 
\end{equation}
for $\left|\mathbf{B}\right|$ in~T. 
So, for example, if
$\left|\mathbf{B}\right| = 0.05$, $n_{15}=1$, $T_6 = 1$, $\beta = 8
\times 10^{-6}$, and the waves are inertial.
The inertial regime pertains 
for substantial distances into the chromosphere (down to about
1500~km above the photosphere in the VAL-C semi-empirical model). 
Note that in other regions of the solar atmosphere, such as in long active region loops with a relatively small magnetic field, the KAW is appropriate, but not in the high magnetic field strength relevant to a flare. 

We have in mind an IAW disturbance with the form of a wave pulse or
simple wave, a case considered  by
\cite{1994JGR....9911095K} and \cite{2007JGRA..11204214W}. However, acceleration
in IAWs is also discussed in the context of global  resonances of the
magnetospheric field
\citep[e.g.,][]{2002JGRA.107gSMP19W,2003JGRA.108cSMP22W,2003GeoRL..30pSSC2W},
which  could be established by repeated partial reflections of the IAW
from the photospheric or low chromospheric boundary. Evidently, the
exact nature of the oscillation will have to be determined in a
self-consistent way along with the particle acceleration. 

\subsection{The electric field strength and electron energy}\label{sect:field}

Described in two-fluid MHD, a large-scale Alfv\'enic perturbation
causes particle cross-field drifts; an $\bf{E}\times\bf{B}$ drift
equal for both species, and a polarization drift.  The ion
polarization drift is a factor $m_i/m_e$ faster than that of the
electrons, constituting a net cross-field current, the magnitude of
which depends on the wave amplitude at a given position.  A
field-aligned current of electrons flows to maintain plasma
quasi-neutrality.

From \cite{2000SSRv...92..423S}, their Equation 47, the relationship
between the perpendicular electric field $E_\perp$ and the change in
the perpendicular magnetic field $b_\perp$ is
\begin{equation}\label{eq:eperp}
E_\perp = v_A b_\perp \left(1+k_\perp^2\lambda_e^2\right)^{1/2},
\end{equation}
which is a modification of the ideal MHD relationship. 
Here $k_\perp = 2\pi/\lambda_\perp$ is the
perpendicular wavenumber of the magnetic disturbance and $\lambda_e$
is the electron skin depth ($=c/\omega_{pe}$; $\omega_{pe}$ being the
electron plasma frequency). 
In Eq.~\ref{eq:eperp} we have also used the fact that the
perpendicular scale of the magnetic disturbance is much larger than
the ion Larmor radius (see also Chaston et al. 2002). 

The relationship between the parallel and perpendicular components of
the wave is given by \cite{2000SSRv...92..423S}, Equation 43:
\begin{equation}\label{eq:epar}
E_\parallel = \left({k_\parallel k_\perp \lambda_e^2 }\over{1+k_\perp^2\lambda_e^2} \right) E_\perp
\end{equation}
where $k_\parallel$ is the parallel wavenumber. While in the magnetosphere, the ratio $ k_\perp^2\lambda_e^2 $ can be comparable to unity, in the solar atmosphere it is typically much less than unity. Therefore the ratio between parallel and perpendicular electric field in the solar atmosphere is going to be small in the solar atmosphere. However, since the perpendicular electric field calculated from Eq.~\ref{eq:eperp} is large, this small fraction can still result in a parallel field large enough to be interesting.
In the absence of precise knowledge about these scales we investigate 
the parameter regimes in which substantial field-aligned electric fields might
be obtained.

To be effective in accelerating electrons, $E_\parallel$ must exceed
the local Dreicer field, $E_D$
\citep{1959PhRv..115..238D,1982SSRv...31..351S}, above which the bulk
of the thermal electron distribution will be freely accelerated
(`runaway'). The Dreicer field, $E_D$, is
\begin{equation}\label{eq:dreicer}
E_D = {{e \ln \Lambda}\over{4\pi\epsilon_o\lambda_{D}}^2} =  {{{e \ln \Lambda}\over{{4\pi\epsilon_o}^2}}
{{n e^2}\over{\epsilon_o k_B T}}},
\end{equation}
where $\ln \Lambda$ is the Coulomb logarithm, $\epsilon_o$ the
permittivity of free space and $\lambda_D$ is the Debye length. $\ln
\Lambda$ is usually taken to be between 20 and 25 for the corona.
In the partially ionized plasma of the lower chromosphere,
$\ln \Lambda$ is modified to $x \ln \Lambda + (1-x)\ln
\Lambda'$ where $x$ is the ionization fraction and $\ln \Lambda'$ the
`effective Coulomb logarithm' describing the interaction of charged
and neutral particles \citep[see e.g.,][]{1973SoPh...28..151B}.

\begin{figure}
\plotone{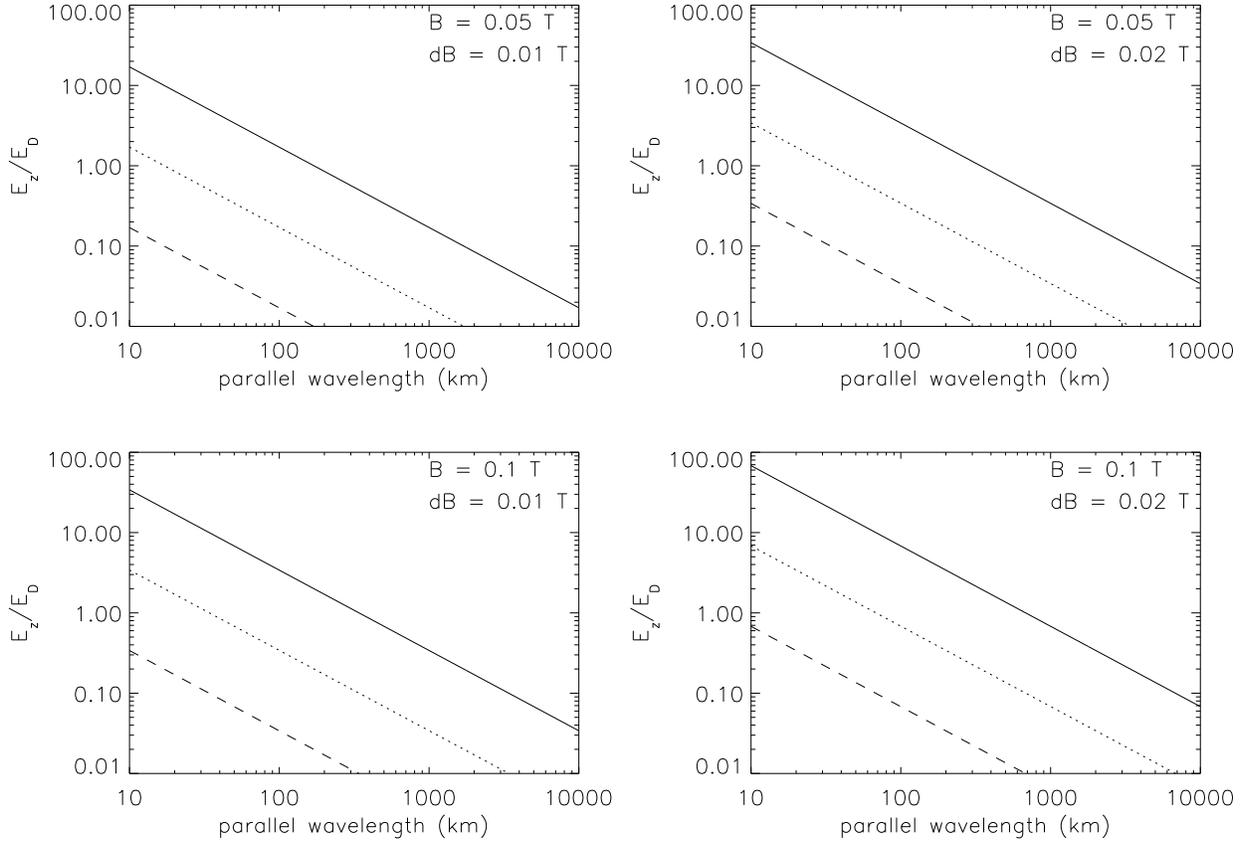}
\caption{The ratio of parallel electric field to Dreicer field for
  field and perturbation values given in the top right corner of each
  panel. The local coronal electron density is $10^{15}{\rm m}^{-3}$
  and the temperature is $10^6$~K. The lines correspond to
  $\lambda_\perp$=0.5~km (solid), 5~km (dotted), 50~km (dashed).}
\label{fig:eratio1}
\end{figure}

\begin{figure}
\plotone{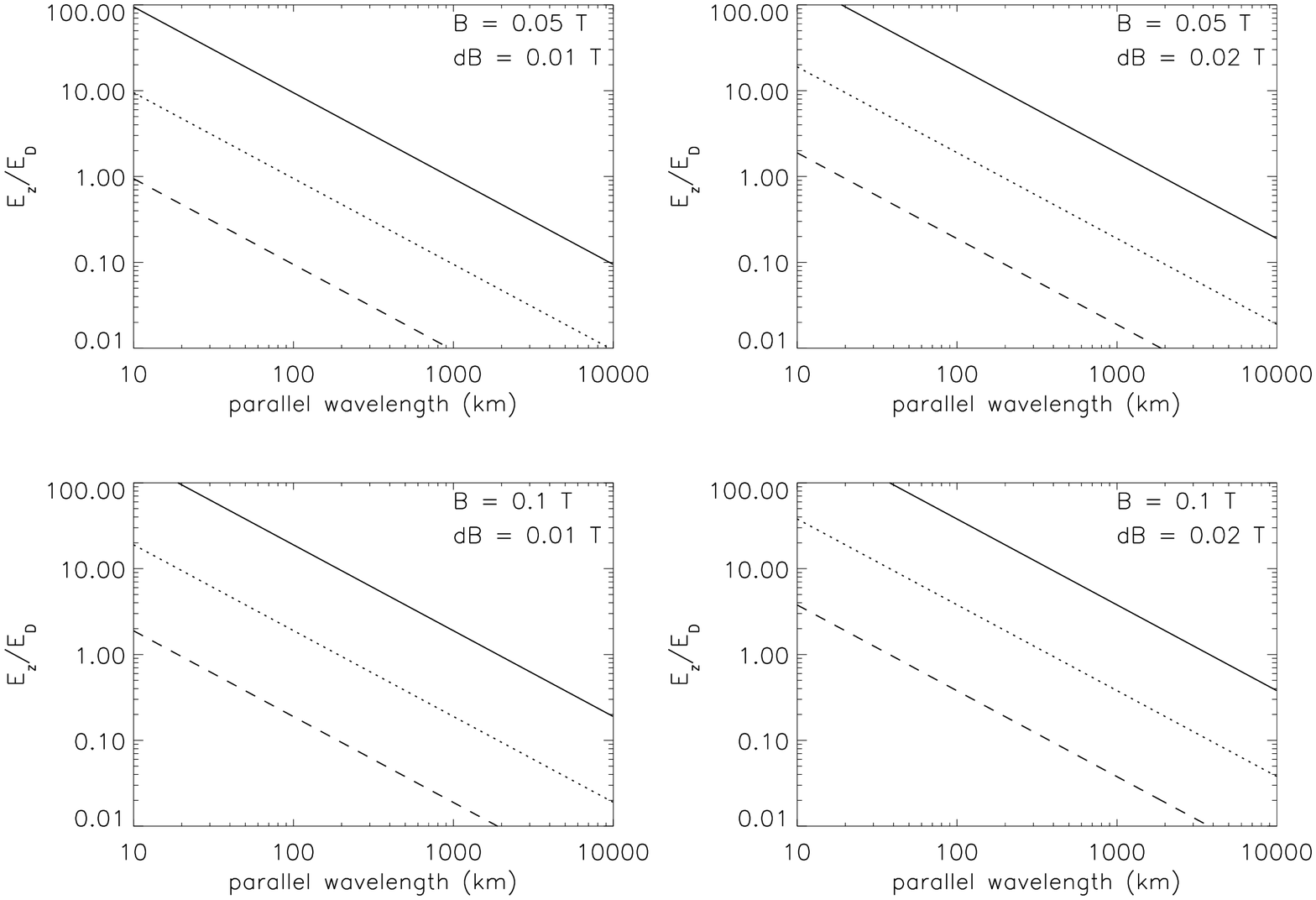}
\caption{As in Figure~\ref{fig:eratio1}, but with an electron density of 
$5\times 10^{14}{\rm m}^{-3}$. }
\label{fig:eratio2}
\end{figure}

Neglecting the temperature-dependence of the Coulomb logarithm, the
ratio of the parallel electric field to the Dreicer field in the
corona is
\begin{equation} {E_\parallel \over E_D} = {10^5{T_6\over
      n_{15}^{5/2}}{\left|\mathbf{B}\right| b_\perp \over{l_\parallel
        l_\perp}}}
\end{equation}
where $l_\parallel,\; l_\perp$ the parallel and perpendicular
wavelengths in kilometers and $\left|\mathbf{B}\right|, b_\perp$ are 
in T. Evidently, only in hot, tenuous, strongly-magnetized plasmas 
will this ratio exceed unity; the ratio is
plotted in Figures~\ref{fig:eratio1} and~\ref{fig:eratio2}, where it
can be seen that at a coronal density of $10^{15}{\rm m^{-3}}$, ${E_\parallel \over E_D} $
exceeds unity only for scales $l_\parallel \sim 10 - 100$~km and
$l_\perp \leq 5$~km. Increasing the temperature, field strength, or
the perturbation amplitude, or decreasing the length-scale of the
perturbation gives a higher value for ${E_\parallel \over E_D} $. However, 
wave-generated super-Dreicer fields are not possible in the chromosphere for 
realistic parameters of the ambient medium or perturbation.

A full calculation of the electron energy spectrum accelerated must be
left for future investigations, as it requires a simulation capable of
following the non-linear evolution of the wave and of the electron
distribution function
\citep[e.g.,][]{2004PhPl...11.1277W,2005JGRA..11001201D}.  But we can
observe that, in a corona of density $5\times 10^{14}{\rm m}^{-3}$,
super-Dreicer fields are produced in strong fields, by propagating
wave pulses having parallel wavelengths of around 100~km and
perpendicular wavelengths of around 5~km. Electrons with a thermal
speed similar to the wave phase speed can be accelerated, via a single
interaction with the traveling wave front, up to twice the Alfv\'en
speed \citep{chaston06} thus gaining 4 times the `Alfv\'en energy',
${1\over 2} m_e v_A^2$, corresponding to 27~keV for $B = 0.05$~T, and
$n = 5\times 10^{14}{\rm m}^{-3}$. The maximum instantaneous electron
flux from a single interaction of electrons with the wave field is $n
v \sim 5\times 10^{14}\ {\rm m}^{-3} \times 2 \times 4.9 \times
10^7{\rm m s^{-1}} = 4.9 \times 10^{22}{\rm m^{-2} s^{-1}}$ (this is
comparable with typical electron fluxes inferred from hard X-rays of
$10^{36}\ {\rm s}^{-1}$ over an area of perhaps $10^{13} - 10^{14}\
{\rm m^{2}}$). However, this flux will only be achieved if all
electrons are accelerated, which will not happen because of the
required velocity resonance condition of the electrons with the
wavefront. Thus, waves with scales of tens to hundreds of kilometers
may be capable of providing a modest flux of coronal electrons at 10
-- 30~keV, running ahead of the wave front.

\subsection{First-order Fermi acceleration in a moving mirror}~\label{sect:fermi1}

Further acceleration can occur via a first-order Fermi process as the
Alfv\'enic wave front, itself a moving mirror, approaches a magnetic
mirror in the lower corona and chromosphere. In repeated reflections,
the parallel electron speed would be increased by $2v_A$ at each
interaction, until the  resulting decrease of pitch angle allows the
electron to penetrate  the mirror. There
is thus the possibility to accelerate a fraction of the injected
electrons up to significantly higher energies.

For repeated reflections, the mirroring electrons must not be
collisionally stopped between one interaction with the wavefront and
the next. So the separation in column depth between wave-front and
mirror must be less than half of the collisional stopping column depth
of the electrons at 2$v_A$ (neglecting the decreasing distance between
wave-front and mirror as the pulse approaches the
chromosphere). Using the expression from \cite{1978ApJ...224..241E},
the collisional stopping column depth of an electron of energy $E$ (in
keV) is:
\begin{equation}
N_c = 10^{21}\mu_e E^2{\;\rm m^{-2}}.
\end{equation}
where $\mu_e$ is the electron pitch-angle cosine. So for an electron
at 20~keV (i.e. following its first encounter with the wavefront),
with a pitch angle of 45$^o$, $N = \int n dl = 2.8\times 10^{23}{\rm
  m^{-2}}$. The electron must therefore mirror within $1.4\times 10^{23}{\rm
  m^{-2}}$. An underdense corona, of $n < 10^{15}{
\  \rm m^{-2}}$ with a loop half-length of $10^7$~m, has $N < 10^{22}\ {\rm
  m^{-2}}$, so a 20~keV electron could penetrate some way into the
chromosphere - to a depth of around 1700~km above the photosphere in 
the VAL-C chromospheric model (a column mass of
$2.3\times 10^{-4}\ {\rm kg\;m^{-2}}$). Thus, an electron could cross
the corona and chromosphere, mirror quite deep down and return for further
acceleration -- producing bremsstrahlung emission en route. The
details of this should be worked out in future.

\subsection{Turbulent Acceleration and Heating in the
Chromosphere}\label{sect:turb}  
We have seen that - with the possible exception of electron acceleration in their parallel electric field - Alfv\'en wave pulses will not dissipate significantly in the corona. This 
leads us to consider the consequences when the wave reaches the chromosphere, and to discuss ways in which the wave energy could be damped there. It is well known that, in a strongly-magnetized atmosphere,  it is not easy to damp Alfv\'en waves by straightforward collisional means, either by ion-electron (Joule) or ion-ion (viscous) collisions \citep[e.g.][]{1961ApJ...134..347O}. For this reason the dissipation of wave energy is normally thought to happen via a cascade process, with the energy ending up in wavelengths small enough for the Joule and viscous processes to be significant. (Note, ion-neutral damping probably is significant in the chromosphere and we return to this later). If such a cascade can develop, it will result in chromospheric heating, but possibly also electron acceleration. The
theory of stochastic electron  acceleration
\citep[e.g.][]{1992ApJ...398..350H,1994ApJ...425..856L,1996ApJ...461..445M,1997ApJ...482..774P,2004ApJ...610..550P,2004ApJ...614..757Y,2006ApJ...644..603P}
provides a possible mechanism for the acceleration of electrons into a
broad spectrum extending to the high energies that are observed. There
is an extensive literature on such acceleration processes; the reader
is directed to  \cite{2002SSRv1011A}  (Section 5.2) for an
overview of the process. Here we will mention only some aspects pertinent to the
application of ideas of stochastic  electron acceleration in this wave model
within the collisional environment of the chromosphere. Firstly we discuss briefly the generation of the cascade itself.

As discussed in Section~\ref{sect:mechanism} it is reasonable to
expect that some fraction of the Alfv\'en mode energy that arrives at
the chromosphere will be reflected at the steep gradients within the
chromosphere or from the  photosphere, and allow the development of a
turbulent spectrum in the counter-streaming wave field, with fast-mode
and Alfv\'en components. To be  viable,  this should happen quickly -
in less than the wave crossing-time of the chromosphere. There is a
vast literature on the development of magnetic  turbulence, but
\cite{2004ApJ...614..757Y} provide useful expressions for  the
relevant timescales. The Alfv\'en spectrum develops within the
turnover time of the  longest wavelength present, $\lambda_{max}$,
i.e.  $t =\lambda_{max}/\delta v$  \citep[see
also][]{1996ApJ...461..445M} where  $\delta v/v_A = b_\perp/B$,
$\delta v$ being the velocity perturbation.  So a (perpendicular)
cascade with energy injected at wavelengths less than  $\lambda =
(b_\perp/B)$ times the height of the chromosphere can develop as  the
Alfv\'en waves cross the chromosphere. The development of the
(isotropic)  fast-mode spectrum, driven by reflected fast mode waves
\citep{1994A&A...286..275G} or fed  by the Alfv\'en spectrum, depends
on the plasma $\beta$.  In the  high-$\beta$ medium of the low
chromosphere it develops in approximately
$t={(\lambda_{max}/v_A)(v_A/\delta_V)^2}$, so that only energy
injected at relatively short wavelengths will cascade quickly
enough. In the low-$\beta$  upper chromosphere the development is yet
slower.  

\begin{figure}
\plotone{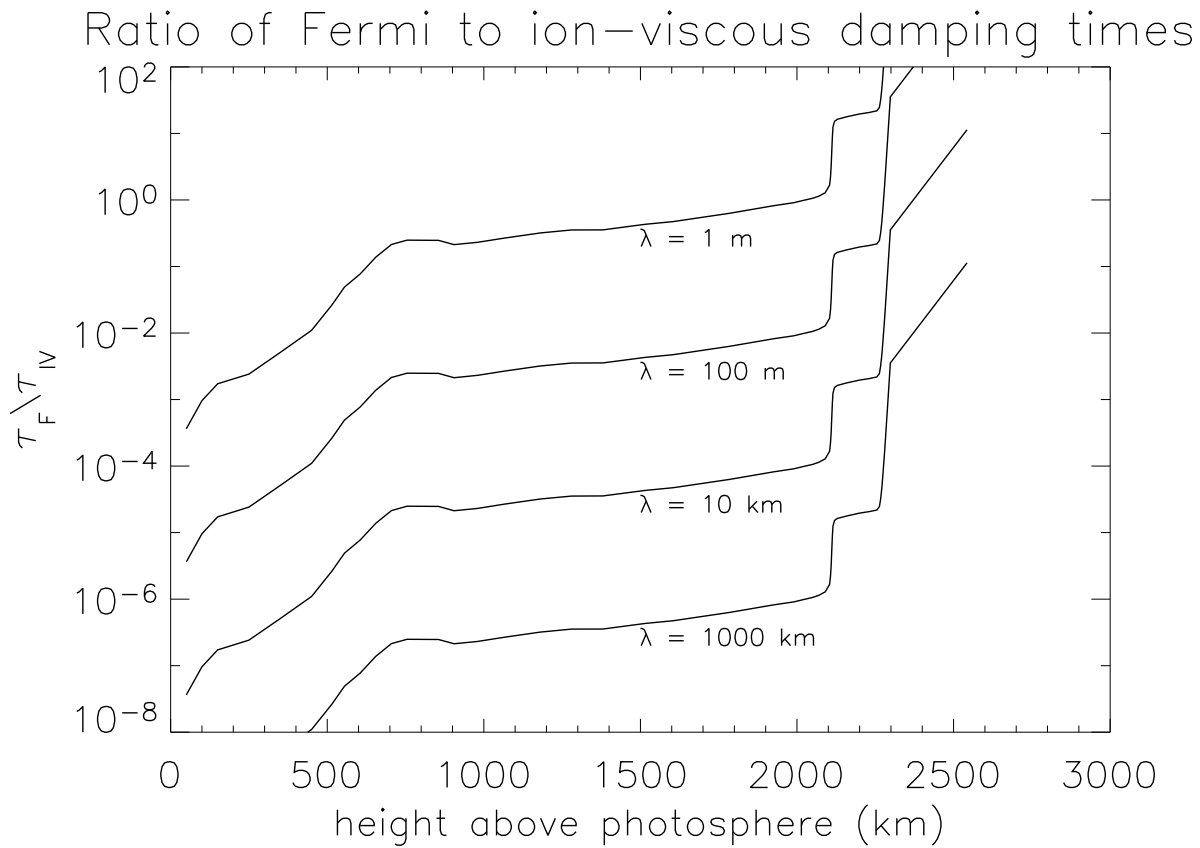}
\caption{Ratio of the Fermi and ion-viscous damping times in the
  chromosphere, using a VAL-C model atmosphere. While this ratio may
  be much larger than unity in the corona, implying that ion-viscous
  damping dominates, it is less than unity throughout the
  chromosphere.}
\label{fig:tauftauiv}
\end{figure}

Damping by Fermi acceleration will dominate in the chromosphere, compared
to ion-viscous damping which may be significant in the corona. We demonstrate this by considering the
ratio of the ion-viscous damping rate to the Fermi damping rate, given by \citep[e.g.][]{2000SoPh..194..131T}:
\begin{equation}
{{\gamma_v}\over{\gamma_F}} ={{\tau_F}\over{\tau_{v}}} = 6 \times
10^{11} {{kT_i^{5/2}}\over{vn_a}}
\end{equation} (converted into S.I. units) where $k$~is the
wavenumber, $T_i$~the ion temperature, $v$~the velocity, and $n_a$~the
density of particles accelerated by the Fermi mechanism. This ratio,
plotted in Figure~\ref{fig:tauftauiv} for a range of different
wavelengths of magnetoacoustic waves, is much less than one in the low
temperature chromosphere, primarily because of the strong temperature
dependence of the ion-viscous damping time.  Therefore,  chromosphere wavelengths
longer than about 1 meter will be preferentially damped by Fermi
acceleration \citep[see also][]{2006ApJ...644..603P} (however in the
corona ion-viscous damping, though weak, can still be dominant). 

Electron acceleration by a turbulent wave spectrum  has been mostly studied
in two main cases; `transit-time' acceleration by low-frequency fast
mode waves
\citep[e.g.][]{1997ApJ...491..939M,1998ApJ...493..451L,2002PhRvL..89B1102Y},
and gyroresonant interaction with a whistler spectrum - the high
frequency end of the Alfv\'en spectrum with $\omega > \Omega_i$
\citep[e.g.][]{1987SoPh..113..195M,1992ApJ...398..350H,2002PhRvL..89B1102Y,2004ApJ...610..550P}.
Of particular importance to us is the effect in these models of
Coulomb collisions: the dense chromosphere might be thought of as
unfavorable for any particle acceleration to exist since energy gained
can be quickly lost again.  Some modeling has considered Coulomb
energy losses and isotropization \citep[e.g.,][]{1992ApJ...398..350H,
1998ApJ...493..451L,2004ApJ...614..757Y}.  In general, one finds that
below the electron energy at which the acceleration timescale exceeds
the collisional loss timescale, the electron distribution is
quasi-thermal. Above this critical energy the distribution can have a
non-thermal character as the electrons become increasingly
collisionless at high energy. For whistler-mode acceleration, the
critical energy is
\begin{equation}
  E_c \sim 3.8{n_{16}}^{3/2}\left({0.01T\over {\left|\mathbf{B}\right|}}\right)^2
  \left({10^{-4}\over R} \right){\rm keV}
\end{equation}
where $R$ is the ratio of turbulent magnetic energy density to total
magnetic energy density \citep[][Eq. 20]{1992ApJ...398..350H}. For
chromospheric parameters of $n_{16} = 100,\; \left|\mathbf{B}\right| =
0.05$ then $E_c = 0.015/R$~keV. If the turbulent energy density fraction
contained in whistlers is $R \sim 10^{-3}$ then the electron
distribution will be non-thermal above 15~keV. It remains to be seen
whether this level of whistler turbulence is plausible.

In the case of transit-time acceleration, which operates at much lower
wave frequencies, \cite{1998ApJ...493..451L} find that energy exchange
between waves and particles is in fact made significantly more
efficient in the presence of Coulomb interactions. This is because
Coulomb collisions (i) exchange energy between accelerated and
non-accelerated electrons, raising the slower electrons up to resonant
energies, and (ii) redistribute the energy gained between parallel and
perpendicular components of momentum, increasing the magnetic moment
of the electrons and thus the rate of the transit-time process.
Transit-time damping by electrons requires that the local electron
thermal speed be comparable to the Alfv\'en speed, equivalent  to
$\beta \sim{m_e}/{m_p}$. Using the VAL-C model, this occurs at around
1500~km above the photosphere, where the density is $\sim 10^{18}\
{\rm m^{-3}}$. It also requires that the wave spectrum be continuous
(as in a turbulent spectrum), or at least have discrete overlapping
modes to allow electrons to stay in resonance as they accelerate.

It should be noted that the simulations of \cite{1998ApJ...493..451L}
are done for a temperature of $3\times 10^6$~K and a density of
$10^{16}\ {\rm m^{-2}}$. It remains to be seen whether the beneficial
trade-off between energy loss and scattering will occur at higher
densities, though since both scattering and loss terms in the
Fokker-Planck equation describing the evolution of the particle
distribution function have the same density dependence (see e.g.
Lenters \& Miller 1998 Eq. 4) we expect that it will.

However, even with the enhanced efficiency provided by Coulomb
collisions, transit-time acceleration does not yield a power-law
distribution as is observed from hard X-rays -- instead it produces
`bulk heating' of electrons, albeit to energies of 10s of keV. Conceivably, 
a low level of whistler turbulence could provide the necessary pitch-angle 
scattering (but without energy redistribution) leading to the formation of an
accelerated non-thermal tail.

\section{Overall Energetics and Open Questions}\label{sect:energetics}
We have introduced the idea of impulsive-phase transport of flare energy 
from its initial site of energy release via Alfv{\'e}n wave pulses,
and in the previous section have shown how this may
lead to the electron acceleration needed to explain the
hard X-ray observations.   A complete theory should also address the
generation of the wave energy in the first place, discuss the
efficiency of the conversion, and describe the regulation mechanisms
that allow the hard X-ray signatures to be so universal.

The partition of energy at its original source poses the first
important question: what fraction goes into the Alfv{\' e}n mode and
what fraction goes into other wave modes?  Emslie and Sturrock (1982)
deal with this question qualitatively and suppose that half of the
energy winds up in the Alfv{\' e}n mode and the other half in the fast
mode, with the slow mode getting a negligible amount because of the
mismatch between the sound speed and the Alfv{\' e}n speed.  To obtain
a better understanding of this energy partition would require a full
understanding of the non-linear development of the energy release,
thus determining the flow fields involved in the deformation of the
magnetic field.   In a low-$\beta$ plasma one would expect this
deformation to proceed at or near the Alfv{\' e}n speed.

The next step in the flow of energy consists of the Poynting flux $S$
of the resulting waves, with $S \sim v_A \times b_\perp^2 / \mu_o$.
The magnitude of the wave field $b_\perp$ can be crudely estimated
from the requirement that this Poynting flux supply the flare energy.
\cite{2007ApJ...656.1187F} show that the broadband flare output in
moderate white-light and UV events, occurring in small footpoint
areas, corresponds to an energy input in excess of $ S \sim 10^{7}\
{\rm J m^{-2} s^{-1}}$.  For an X-class flare energy dissipation of
$10^{25}$ Joules in $10^3$ seconds, over a spatial footpoint scale of
$(10^4~\rm{km})^2$, we need $S \sim 10^8 {\rm J\ m^{-2} s^{-1}}$. For
$v_{A} \lesssim 1 \times 10^{4}\;{\rm km\ s^{-1}}$ at the
chromospheric formation depth of the broad-band emission, then $|{\bf
b_\perp}| \gtrsim$~=0.003~T. This is well within the upper limit to
plausible field variations, given by the permanent line-of-sight field
changes observed at the photosphere in large flares.

Other areas of theoretical uncertainty involve the degree of
reflection of the wave energy on the gradients at and below the
transition region, and the related question regarding the growth  rate
of the turbulent cascade.  In the lower atmosphere the Alfv\'en speed
varies over a scale short compared to the wavelength of the
disturbance, so the disturbance will be partially reflected and
partially transmitted (though the fact that stepwise photospheric
field changes of order ten percent are seen suggests that a
considerable fraction of wave energy is transmitted to the
photospheric level).   Emslie \& Sturrock (1982) discuss wave
transport and dissipation in the context of a normal solar atmospheric
model, in which thermal conduction creates a sharp transition
layer. In this case substantial wave reflection will occur, launching a 
propagating wave towards the conjugate footpoint. The a
coefficient of reflection is given by $R_{E} =
{{(\theta}^{1/2}-1)^{2}/{(\theta^{1/2} + 1)^{2}}}$ where  $\theta$
represents the temperature ratio between corona and chromosphere.  For
a quiet solar atmosphere we might have $\theta$ = 200 and $R_{E} \sim
$75\%, but clearly in a flaring atmosphere this estimate will have to
be modified and will affect the wave energy reaching the chromosphere.  
Strong heating should increase the scale height,  soften the transition region 
and reduce the reflected component. In the radiative hydrodynamic models of \cite{2005ApJ...630..573A}, the density and temperature gradients between chromosphere and corona are indeed 
at first on average smoothed out by atmospheric heating in the impulsive phase, 
but then steeper temperature gradients occur as the corona heats. 
But the behavior also varies with the intensity, and location of heat input, which 
depends of course on the energy transport model and atmospheric structure, and will need to be examined in detail.

The energy of the transmitted fraction will be dissipated in the
chromosphere. Alfv\'enic
disturbances can damp resistively, if on small enough scales, or by
other means such as ion-neutral coupling which may be particularly
important in the lower chromosphere. De Pontieu et al. (2001)
considered the damping by ion-neutral coupling in the lower
chromosphere of large-scale coronal oscillations, observed in TRACE to
be excited by flares and filament eruptions \citep{2002SoPh..206...69S}.
Although these waves are kink (fast mode) waves in flux tubes with
relatively low fields, analogous damping may occur for our Alfv{\' e}n
mode waves in strong field regions. The Joule dissipation as
calculated by Emslie \& Sturrock is enhanced by a factor $(1+s)$ where
$s$ is the ``ion slip'' term;
\begin{equation}
 s = \left({\rho_n\over
    \rho_t}\right)^2 {{\Omega_e \Omega_i}\over{\nu_{\it eff}
    \nu_{in}}};
\end{equation}
here $\Omega_e$ and $\Omega_i$ are the electron and ion
gyrofrequencies, $\nu_{\it eff} = \nu_{ei}+\nu_{en}$, the collision
frequencies of electrons on ions and neutrals, respectively, and
$\nu_{in}$ is the ion-neutral collision frequency. De Pontieu et al.
found the slip $s$ to be large throughout the chromosphere, resulting
in Joule heating that peaks between around 300~km and 1000~km above
the photosphere. This is close to the temperature minimum region where
localized energy input is required to generate the observed
white-light flare continuum excess.

Finally, any remaining undamped waves, once reflected at the photosphere 
or at strong chromospheric gradients may lead to the development of a turbulent 
cascade which, as we have noted, provides one of the major
possibilities for chromospheric electron acceleration. Again a quantitative
description of this partitioning is beyond the scope of this paper.

\section{Conclusions}
 
Energy transport by Alfv{\' e}n waves has a well-developed literature
in the context of the terrestrial aurora, and we have applied similar
ideas here to the problem of flare effects in the solar atmosphere.
Our new understanding of active-region magnetic fields, based on
microwave observations, now  convinces us that the transport time for
these waves is very short -- short enough to explain the rapid time
variations and tight conjugacy of double-footpoint hard X-ray sources
-- and also that the energy flux can be very large.  From this point
of view, Alfv{\' e}n waves therefore provide an alternative to energy
transport by electron beams. Emslie \& Sturrock aimed at explaining a
relatively weak warming of the temperature-minimum region late in the
flare, as required by Ca~K line observations of Machado et
al. (1978). We instead wish to explain the entire energy of the flare
impulsive phase in this manner.

Replacing the electron beam of the standard thick-target model with an
Alfv{\' e}n-wave Poynting flux implies particle acceleration in the
chromosphere or at the base of the coronal loop carrying the wave.
Because of the dominance of fast electrons in the flare energy budget,
we have discussed mechanisms for electron acceleration in this
scenario at length.  Our analysis establishes the feasibility of these
ideas without pinpointing which of the possible acceleration modes
dominates. 

Finally, we note that the ideas we present are novel in the solar
context but are well-established in the Earth's magnetosphere.  These
ideas should be considered not only for solar flares, but  elsewhere
in the Universe where magnetic reconnection is invoked.

\acknowledgments

We would like to acknowledge many useful discussions with numerous
colleagues, including J.~C.~Brown, P.~Cargill, C.~Chaston, P.~Damiano,
N.~Gopalswamy, G.~Haerendel, J.~I.~Khan, E.~P. Kontar,
A.~L.~MacKinnon, J.~P.~McFadden and K.~Shibata. We would also like to
thank an anonymous referee for vigorous discussion which led to
significant improvements in the overall structure and clarity of the
paper. This work was supported by NASA under grants NNG05GG17G
(L. F. and H. S. H) and NAS 5-98033 (H. S. H.), and by PPARC under
Rolling Grant PP/C000234/1 (L.F.).  Financial support by the European
Commission through the SOLAIRE Network (MTRN-CT-2006-035484) is
gratefully acknowledged. 

\bibliographystyle{apj}
\bibliography{ms}

\end{document}